\newcommand{\CLASSINPUTbottomtextmargin}{1.9cm}
\documentclass[conference]{IEEEtran}

\usepackage{cite}
\usepackage{url}
\usepackage{amsmath,amssymb,amsfonts}
\usepackage{algorithmic}
\usepackage{graphicx}
\usepackage{textcomp}
\usepackage{multirow}
\usepackage{xcolor}
\usepackage{booktabs}
\usepackage{enumitem}
\usepackage{comment}
\usepackage{etoolbox}


\def\BibTeX{{\rm B\kern-.05em{\sc i\kern-.025em b}\kern-.08em
    T\kern-.1667em\lower.7ex\hbox{E}\kern-.125emX}}
\begin{document}

\title{In-place Double Stimulus Methodology for Subjective Assessment of High Quality Images\\
}

\author{
\IEEEauthorblockN{
Shima Mohammadi\IEEEauthorrefmark{1},
Mohsen Jenadeleh\IEEEauthorrefmark{2},
Michela Testolina\IEEEauthorrefmark{3},\\
Jon Sneyers\IEEEauthorrefmark{4},
Touradj Ebrahimi\IEEEauthorrefmark{3},
Dietmar Saupe\IEEEauthorrefmark{2},
João Ascenso\IEEEauthorrefmark{1},}
\IEEEauthorblockA{
\small
\IEEEauthorrefmark{1}IST-IT, Portugal \hspace{0em}
\IEEEauthorrefmark{2}University of Konstanz, Germany \hspace{1em}
\IEEEauthorrefmark{4}Cloudinary, Belgium \hspace{1em}
\IEEEauthorrefmark{3}EPFL, Switzerland \hspace{1em}
\IEEEauthorblockA{
\footnotesize
\texttt{\{shima.mohammadi, joao.ascenso\}@lx.it.pt, \{mohsen.jenadeleh, dietmar.saupe\}@uni-konstanz.de} \\
\texttt{jon@cloudinary.com, \{michela.testolina, touradj.ebrahimi\}@epfl.ch}
}
}
}

\maketitle

\begin{abstract}
This paper introduces a novel double stimulus subjective assessment methodology for the evaluation of high-quality images to address the limitations of existing protocols in detecting subtle perceptual differences. The In-place Double Stimulus Quality Scale (IDSQS) allows subjects to alternately view a reference and a distorted image at the same spatial location, facilitating a more intuitive detection of differences in quality, especially at high to visually lossless quality levels. A large-scale crowdsourcing study employing this methodology was conducted, generating a comprehensive public dataset to evaluate perceived image quality across several compression algorithms and distortion levels. An additional contribution is the modeling of quality scores using a Beta distribution, allowing for the assessment of variability and subject consistency. Our findings demonstrate the effectiveness of the IDSQS methodology in achieving high correlation with more precise subjective evaluation benchmarks. The dataset, subjective data, and graphical user interface developed for this study are publicly available at \url{https://github.com/shimamohammadi/IDSQS}.

\end{abstract}


\section{Introduction}

\par Full-reference image quality assessment is broadly divided into two main approaches: subjective and objective. Objective methods, such as Peak Signal-to-Noise Ratio (PSNR) and Structural Similarity Index Measure (SSIM) \cite{wang2004image}, predict the fidelity of images with respect to their reference. Despite their computational efficiency, these metrics often fail to reflect the perceptual and contextual factors that influence human evaluations of image quality. Subjective methods, which rely on human observers making judgments based on their personal experience, are widely considered the gold standard for assessment of image quality. Despite the efficiency and speed offered by objective metrics, their development, training (in the case of data-driven models), and validation depend on scores obtained from subjective assessment methods.

\par The design and methodology of subjective tests have been standardized by organizations such as the International Telecommunication Union (ITU), with key guidelines outlined in \mbox{ITU-T} Recommendation BT.500 \cite{BT.500} and ITU-R Recommendation P.910 \cite{P.910}. These methodologies are widely used by the research community in many areas, such as image and video compression, super-resolution, and denoising. However, these traditional methodologies face significant limitations in effectively assessing high to visually lossless image qualities, as produced by recent image compression solutions. This shortcoming in the existing evaluation methods has motivated the establishment of a new activity on the Assessment of Image Coding, known as JPEG AIC-3. This activity builds on the previous standardization efforts (AIC-1 and AIC-2) and seeks to \textit{specify standards or best practices w.r.t.\ subjective and objective image quality assessment methodologies, respectively, for a range from high quality images to mathematically lossless images}. As part of this effort, a ranking-based subjective test methodology based on triplets was proposed, evaluated and then adopted. In this case, human subjects rank images according to their perceived quality, enabling a more sensitive detection of artifacts in compressed images.

\par This paper introduces a novel methodology to advance state-of-the-art of direct rating methods, which may not be as accurate as ranking based methods (such as those proposed in JPEG AIC-2/3) but offers a offers a quality scale that is more intuitive, easy to interpret and aligned with natural language descriptions. The proposed double stimulus methodology is assessed for its validity and effectiveness in evaluating high to nearly visually lossless images. Furthermore, its limitations and potential drawbacks are analyzed, with particular focus on their influence on the accuracy and reliability of the quality scores. The primary contributions of the paper are as follows:
\begin{enumerate}[leftmargin=*]
    \item In-place Double Stimulus Quality Scale (IDSQS): This methodology involves showing only one image at a time, rather than displaying the reference and distorted images side-by-side or sequentially. Participants can freely toggle between the reference and the distorted image. In-place comparisons, as opposed to side-by-side or sequential viewing, make it easier for test subjects to identify differences in quality. 
    Also, this methodology allows for the display of high-resolution images since only a single image is viewed at time. Other details of the proposed design, such as the scale used, are discussed in Section \ref{sec:subj_test_meth}.
    \item A large-scale crowdsourcing subjective assessment test using the proposed IDSQS methodology was conducted. Both the results of the test and the graphical user interface developed for this study are made publicly available at \url{https://github.com/shimamohammadi/IDSQS}. 
    \item The processed data was further analyzed using the Beta distribution to model the scores distribution. Additionally, the performance of the proposed method was compared against the more precise triplet comparison methods already established in the literature.
\end{enumerate}

Thus, IDSQS presents reference and test images in the same location via toggling, avoiding spatial bias and alignment issues common in side-by-side or sequential setups. Moreover, it requires only one rating per trial (as opposed to DSCQS), reducing cognitive load and simplifying the task. It is also easier to implement in online or crowdsourced settings, where full control over layout, resolution, or timing is not feasible. In addition, novel methods were used for processing and analysis of the collected data. Trap questions were employed to identify unreliable responses, the consistency of test subjects’ scores was evaluated to ensure coherence and stability, and the soft quality reconstruction method of ITU-T P.910 was also applied.

\section{Related work} \label{sec:related_work}

Standardized protocols for subjective visual quality assessment are predominantly based on the guidelines outlined in ITU-R BT.500, ITU-T P.910 and the ISO/IEC TR 29170 (JPEG AIC). The BT.500 and P.910 recommendations provide guidelines and best practices for conducting subjective image quality evaluations, encompassing both single and double stimulus testing methodologies. Among these, the Double Stimulus Continuous Quality Scale (DSCQS) protocol is widely utilized. This method involves subjects evaluating the quality of two stimuli displayed simultaneously (side by side) on a continuous scale ranging from ``bad'' to ``excellent''. The reference stimulus is randomly positioned, and subjects are not explicitly informed of its presence. The DSCQS protocol is particularly effective for assessing compression techniques that enhance visual appeal or target quality levels ranging from low to medium-high. Another frequently applied methodology, the Double Stimulus Impairment Scale (DSIS), requires test subjects to rate the distorted image relative to a reference using a five-point categorical scale, offering a straightforward approach for assessing impairments. Single stimulus methods such as Absolute Category Rating (ACR) method are also defined in both BT.500 and P.910. These are direct category judgment methods where the test stimuli are presented once at a time and rated independently on a category scale. 

The JPEG standardization committee has produced two specifications targeting image quality assessment, AIC-1 and AIC-2 \cite{aic1, aic2}. The first focuses on defining common vocabulary and guidelines for subjective, objective, and computational evaluation of image coding systems, including a review of BT.500 \cite{BT.500} and BT.1082 \cite{BT1082}. In contrast, the AIC-2 \cite{aic2} presents two novel methodologies for the assessment of visually lossless image compression, namely AIC-2 Annex A and Annex B. In Annex A, a triplet of two distorted images and their associated reference image is presented, and subjects are asked to select the image that most closely resembles the reference. In Annex B, the flicker technique to enhance detection of subtle differences is used. In this case, the distorted and reference images alternate rapidly with the reference image to create a flickering effect, and subjects are asked to select the one that does not exhibit flickering.

More recently, JPEG has launched a new initiative, AIC-3, which defines standardized methodologies for assessing images of high quality up to mathematically lossless \cite{ISO29170-32025}. 
A comprehensive analysis of the DSCQS and JPEG AIC-2 Annex A and Annex B subjective test methodologies, for this quality range has been published in \cite{testolina2023performance}. The findings reveal that all three methods exhibit low performance and concludes about the need of robust methodologies. Thus, a Call for Contributions was issued by the JPEG committee to define specific techniques to address such limitations. Build on the work proposed in~\cite{men2021subjective}, JPEG AIC-3 standard~\cite{ISO29170-32025} introduces a Boosted Triplet Comparison (BTC) technique to enhance observers' ability to detect artifacts using methods such as zooming, artifact amplification, and flickering. To allow for scaling the obtained scores with BTC to the desired original perceived distortion magnitudes under normal viewing conditions, a smaller number of Plain Triplet Comparisons (PTC) without boosting is included. The subject responses obtained from the two test protocols (BTC-PTC) are then processed with a statistical numerical method to derive the desired unified Just Noticeable Difference (JND) quality scale. This methodology and its associated data processing is described in \cite{testolina2025fine,jenadeleh2025subjective,ISO29170-32025}.

\section{Proposed Subjective Test Methodology}\label{sec:subj_test_meth}
\begin{figure}
    \centering
    \includegraphics[width=0.9\linewidth]{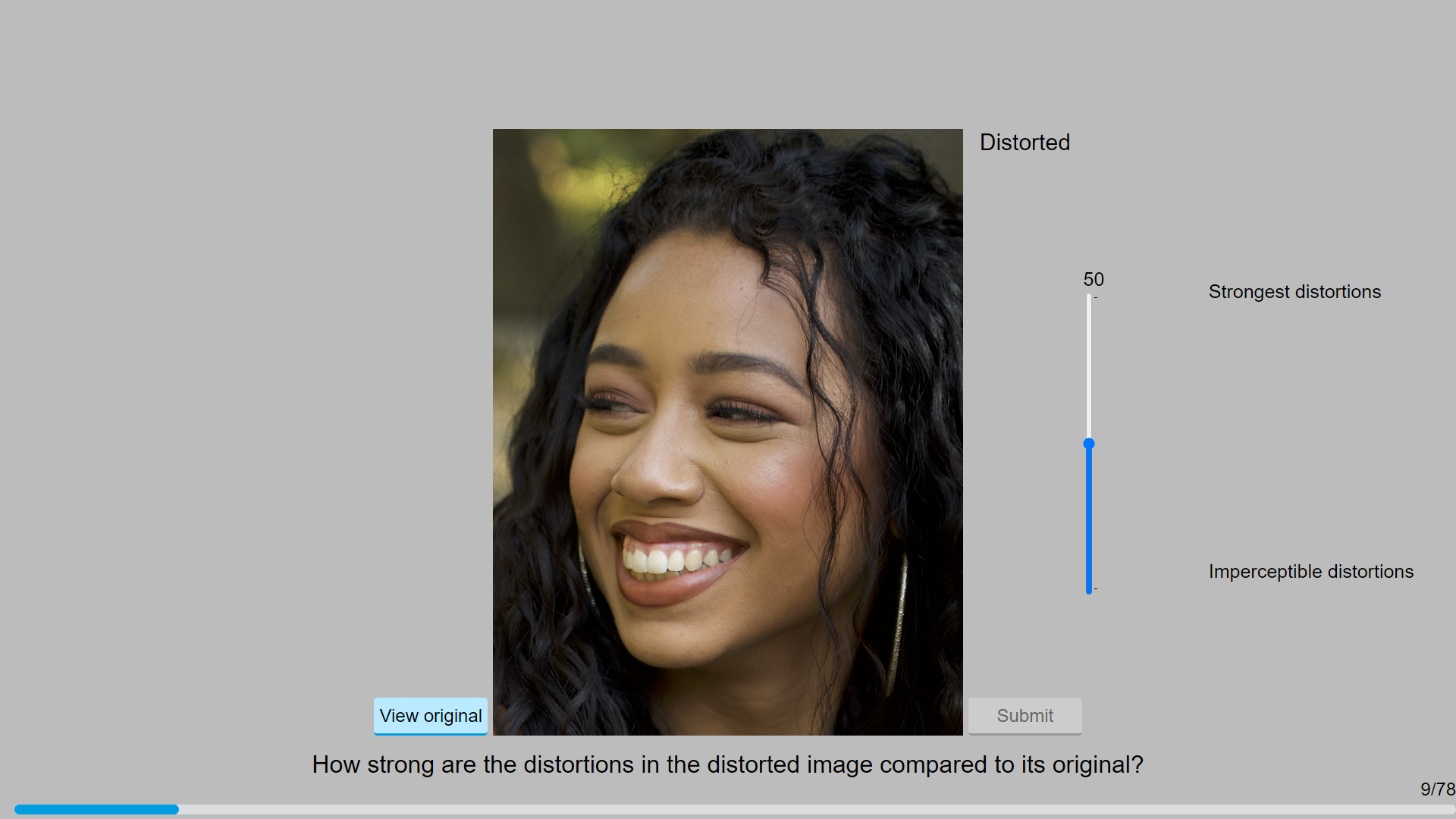}
    \vspace{-8pt}
    \caption{Layout of the interface developed.}
    \label{fig:interface}
\end{figure}

\vspace{-2pt}
The proposed IDSQS subjective test methodology is designed to evaluate image quality through direct comparison between a reference image and the corresponding distorted image. Subjects are presented with pairs of images on a display interface, where one image is the reference/uncompressed image and the other is its distorted/compressed version. Unlike other double stimulus methodologies such as DSCQS, which depend on short-term memory due to sequential image display (with a gray screen in between), or suffer from the rapid eye movement challenges of side-by-side comparisons, IDSQS displays the images at the same location. This allows the subject to detect small changes between the images of a pair and thus make better decisions. Note also that is very different from JPEG AIC-2 since flickering is not applied, neither pairwise selection (ranking based). A layout of the interface that was designed is shown in Fig.\ref{fig:interface}.

Subjects can toggle between the two images by pressing and holding a toggle button. To prevent flickering, a maximum of two toggles per second is enforced. Flickering is considered a boosting technique that amplifies the visibility of distortions in distorted images, which is undesirable in this case, as it could bias the evaluation of image quality. Subjects are instructed to rate the degree of impairment in the distorted image relative to the reference on a continuous scale ranging from 0 to 100, where 0 represents the highest quality (indistinguishable from the reference) and 100 indicates the lowest quality (severe distortion). This continuous scale allows a fine-grained differentiation in perceived quality compared to other methodologies such as in DSIS where just a few levels can be selected or in DSCQS, where the scale includes level markings, thereby influencing the subject decision. The aim of these changes is to offer a finer granularity which is rather important to differentiate images in the high to visually lossless qualities.

\vspace{-5pt}
\section{Experimental Setup}\label{sec:setup}

This section describes the test material, test questions, and the crowdsourcing campaign.

\subsection{Test Material}

Five images from the JPEG AIC-3 dataset \cite{testolina2023jpeg} were selected (of the 10 available) to represent a diverse range of image types and content. These specific images were also previously utilized in the development of the BTC-PTC methodologies for the upcoming AIC-3 standard enabling a direct comparison. The dataset was obtained using five lossy compression techniques, JPEG, JPEG 2000, AVIF, VVC Intra, and JPEG XL, at ten different bitrates, which represent distortion levels ranging from 1 to 10, with 10 corresponding to the lowest bitrate and quality. These distortion levels are spaced approximately 2.5 JND units for each source and content, covering a quality spectrum from high to nearly visually lossless. For subjective assessment, images were cropped to $620\times800$ pixels in size. 

\subsection{Batches of Study and Trap Questions}
Two types of questions were used in this study:
\begin{itemize}[leftmargin=*]
    \item Study questions: Comprised of image pairs, where one image serves as the reference and the other is a compressed version. The latter is intended to be the primary focus for quality evaluation.
    \item Trap questions: These questions were designed to evaluate the reliability of the responses, still using the reference as one image of the pair. They have two sub-categories: 
    \begin{itemize}[leftmargin=*]
        \item Type I: Trap questions where the degraded image has a distortion level of 10. Thus, they exhibit the highest distortion compared to the reference image; subjects are expected to clearly perceive the distortions and rate the images close to a score of 100.
        \item Type II: Trap questions where the degraded image has a distortion level of zero. In this case, no difference between the reference image and the distorted image exists; subjects are expected assign a score close to zero.
    \end{itemize}
\end{itemize}
The study and trap questions were randomly distributed across four batches. Each batch consisted of 79 study questions (each including a distorted image and its corresponding reference) and 10 trap questions. The questions were presented to participants in a randomized order to minimize potential bias.

\subsection{Crowdsourcing Test}
Subjects were recruited through Amazon Mechanical Turk (MTurk), with each subject restricted to completing a single assignment. Each assignment included two randomly drawn batches. After completing the work on the first batch, subjects could opt to do the second batch, after a mandatory three-minute break. Before the start of the tests, comprehensive instructions were provided to subjects, followed by a visual acuity test, a mandatory consent form, and a training phase. The visual acuity test used plates 3 and 4 of the Ishihara color vision test \cite{naderi2024crowdsourcing}. The training phase included both easy-to-answer and difficult questions, during which subjects were guided on their responses and became familiar with the platform. A minimum screen resolution of 1920x1080 pixels was required, and only PC and laptops were accepted. 

\par For each question, 45 ratings were collected. A total number of 132 subjects completed 179 batch instances. Test subjects were divided into different categories, with the age group 31 to 35 years old being the most common category. Of the total participants, 87 were male and 45 were female. Regarding the display size, the majority of test subjects conducted the experiment with displays of sizes 13’’, 22’’, and 32’’. All experimental procedures were approved by the Institutional Review Board of the University of Konstanz.

\section{Proposed Data Processing}
\label{sec:data_processing}
Three data processing steps were applied in the following order: i) data cleansing (Section \ref{sec:data_cleansing}), ii) outlier removal (Section \ref{sec:outlier_removal}), and iii) Mean opinion score (MOS) quality reconstruction (Section \ref{sec:qual_recon}). 

\subsection{Data Cleansing}\label{sec:data_cleansing}

Since subjects could perform one or two batches with at least a 3-minute break, their behavior might vary between batches. Thus, data cleansing and outlier removal were conducted on a batch-by-batch basis rather than by subject. 

\begin{figure}
    \centering
    \includegraphics[width=0.9\linewidth]{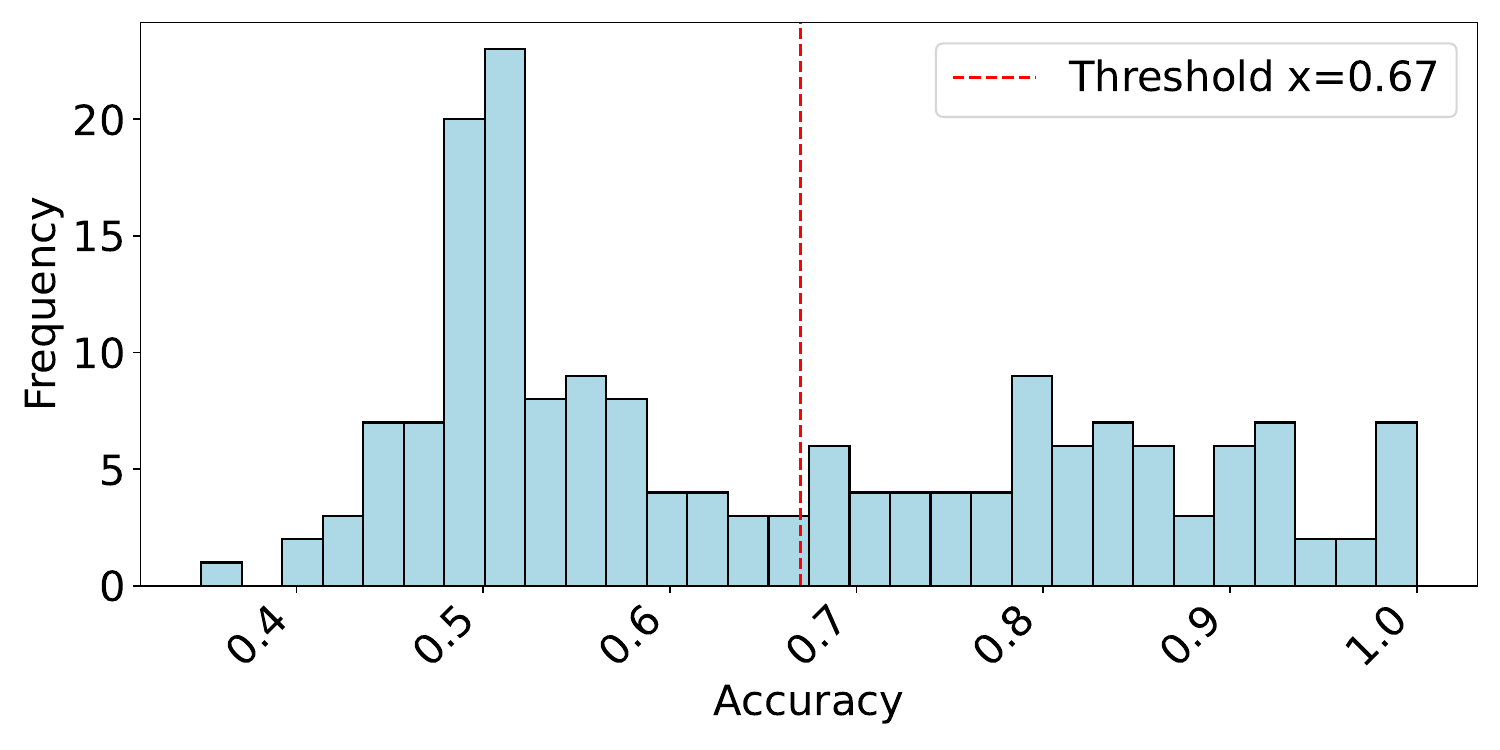}
    \vspace{-15pt}
    \caption{Accuracies for trap questions per batch with Otsu threshold.}
  \vspace{-10pt}

    \label{fig:acc}
\end{figure}

Batches were judged based on the accuracy of the responses to their trap questions. The accuracy of a response $v \in [0,100]$ for a trap question of Type~I is taken to be $v/100$. For a Type~II trap question, it is $1-v/100$. The overall accuracy of a batch is the mean accuracy for the responses to all of its trap questions. The histogram in Fig.~\ref{fig:acc} clearly shows a cluster of batches centered around 0.5 accuracy. This is the accuracy value expected from a random clicker, thereby motivating the use of Otsu thresholding \cite{Otsu}, a method particularly effective for binary classification --- such as separating reliable from unreliable batches as happens in this case. Thus, of the total of 179 batches, 104 with accuracy below the obtained Otsu threshold of 0.67 were discarded as unreliable, which is expected considering the crowdsourcing nature of the test and the difficulty in rating high-quality images.

\subsection{Outlier Removal} \label{sec:outlier_removal}

\par In outlier removal, batch instances are excluded based on the consistency of the scores in a batch, relative to the mean scores across all the batches. Unlike the approach described in the previous section, this method does not depend on trap questions. To enhance the detection and elimination of outlier batches, the newly introduced correlation-based BT.500 recommendation method \cite{BT.500}, published in 2023, was employed. This method was selected for its simplicity and because it does not assume any specific distribution (e.g., Gaussian), unlike the commonly used kurtosis-based approach.

\par The consistency of scores is defined by the minimum value of the Pearson Linear Correlation Coefficient (PLCC) and the Spearman Rank Order Correlation Coefficient (SROCC). In this work, these metrics are calculated at the batch level. Specifically, the minimum correlation $CR_b$ of a batch instance $b$ is determined by the scores for questions $q=1,\ldots,Q$ in a batch $b$, $(S_b(1), \ldots, S_b(Q))$, and the corresponding mean opinion scores, $(MOS_b(1), \ldots, MOS_b(Q))$.

\par Next, the correlation value of a batch, $CR_b$, is compared to values of all correlations, $CR$. A batch $b$ is removed if the $CR_b$ is less than $\min((\mu(CR)-\sigma(CR)), 0.85)$ as in \cite{BT.500}. Based on this approach, 12 batch instances were identified as outliers, leaving 63 batch instances in the final dataset.

\vspace{-5pt}
\subsection{DMOS Quality Reconstruction} \label{sec:qual_recon}

In crowdsourcing environments with difficult to control conditions, it is recommended to explicitly model each subject’s behavior, focusing on factors such as bias (\ref{equ:bias}) and consistency (\ref{equ:inconsist}) that influence scoring. In this study, the ``true quality'' estimate for each question, as outlined in \cite{P.910}, was calculated. This technique consists of an iterative procedure that attempts to jointly estimate the quality of each stimulus together with the bias and consistency of each subject. The estimate represents a bias-corrected, consistency-weighted MOS (\ref{equ:mos_new}). This technique is often referred as ``soft rejection'', as it does not discard any ratings of subjects, but instead assigns lower weights to responses from inconsistent subjects. 

The method starts by initializing the mean opinion score $MOS^0(q)$ for question $q$ from the raw data. In each iteration $n=1,2,\ldots$, it estimates bias (\ref{equ:bias}), inconsistency (\ref{equ:inconsist}), and the MOS (\ref{equ:mos_new}) as follows.
    \vspace{-5pt}
\begin{equation}
    B_i^n = \frac{1}{Q_i}\sum_{q=1}^{Q_i} [S_i(q) - MOS^{n-1}(q)],
    \label{equ:bias}
    \vspace{-5pt}
\end{equation}
where $Q_i$ is the number of questions answered by the subject~$i$, $S_i(q)$ is the score of subject~$i$ in the $q$-th question. Then the residual in each observed rating not explained by the MOS and the subject bias is computed as
\begin{equation}
    R_i^n(q) = S_i(q) - MOS^{n-1}(q) - B_i^n.
\label{equ:inconsist}
\end{equation}
The consistency $W_i^n$ for subject $i$ is the inverse variance $1/ \sigma^2(R_i^n)$ of the residuals. Finally, the MOS is updated, 
\begin{equation}
    MOS^{n}(q) = \frac{\sum_{i=1}^{N} W_i^n(S_i(q)-B_i^n)}{\sum_{i=1}^{N}W_i^n}
    \label{equ:mos_new}
\end{equation}
The iteration terminates when 
\begin{equation}
    \sum_{q=1}^{Q} [MOS^n(q) - MOS^{n-1}(q)]^2 < 10^{-6}.
    \label{equ:terminate}
\end{equation}
Then, the MOS for the source image (distortion level of 0) is subtracted from each calculated MOS to derive the Difference Mean Opinion Scores (DMOS). 

\section{Experimental Results}
\label{sec:exp_results}
In this section, the distribution data scores obtained after processing are analyzed and then compared to the more reliable BTC-PTC JPEG AIC-3 methodology \cite{testolina2025fine}.

\subsection{DMOS Data Analysis}

\par
\begin{figure*}
    \centering
    \includegraphics[width=0.9\linewidth]{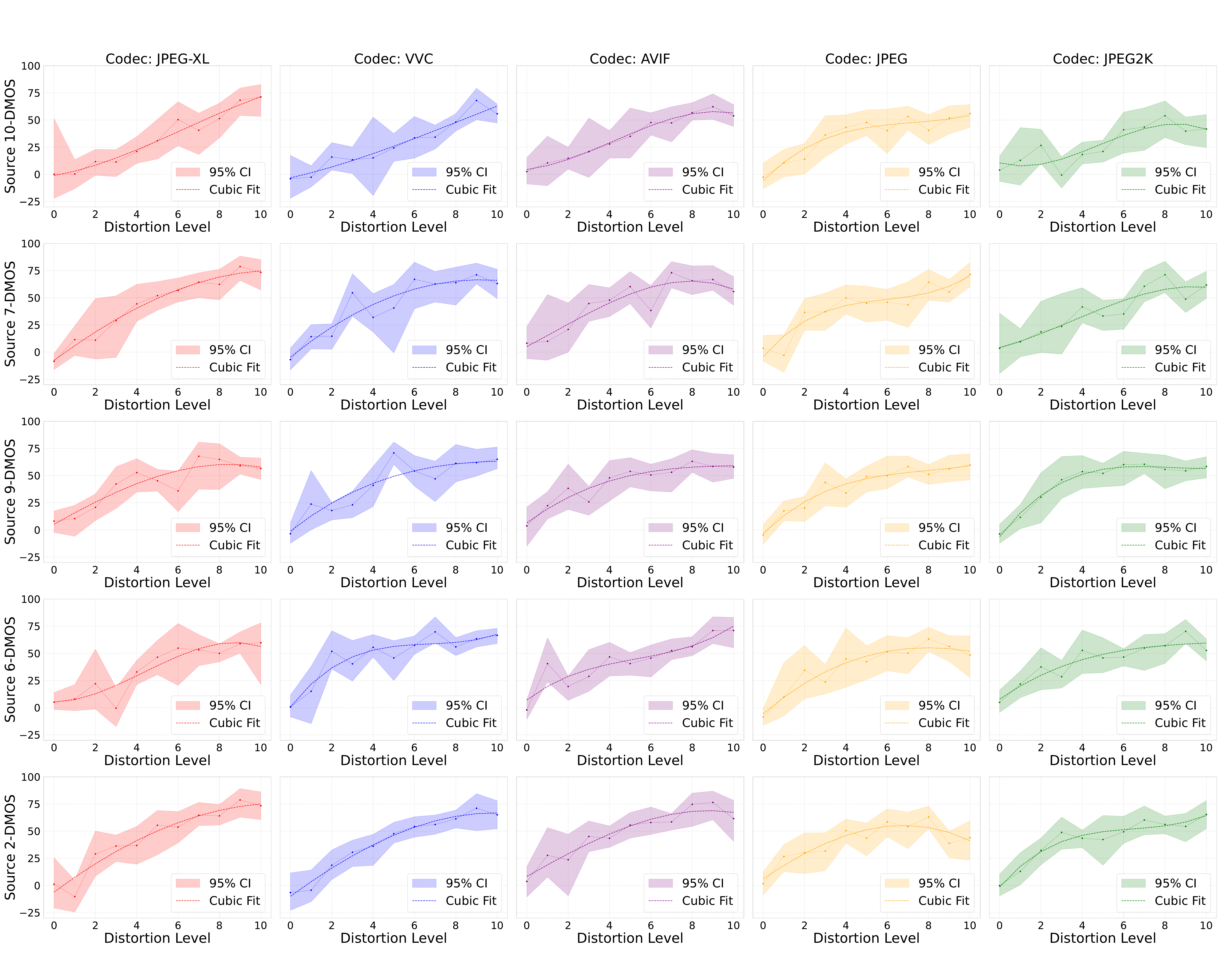}
    \vspace{-15pt}
    \caption{Reconstructed DMOS scale versus distortion levels, where level 0 represents pristine quality (i.e., the source image), and level 10 corresponds to the highest distortion level. Each row shows the plot for one of the five source images, and each column corresponds to one of the five codecs
    }
    \label{fig:recon-scale}
\vspace{-5pt}
\end{figure*}

 A total of $n = 1000$ bootstrap samples were generated by resampling the ratings for each stimulus (with replacement). The scale reconstruction process (Section \ref{sec:qual_recon}) was then repeated for each sample to obtain $n$ values for each MOS. Finally, the $95\%$ confidence intervals (CI) were obtained from the corresponding quantiles. The reconstructed quality scores (points) are shown in Fig. \ref{fig:recon-scale}, along with a fitted cubic curve and the highlighted CIs. The reconstructed qualities should exhibit monotonically increasing curves, reflecting the expected progression of image quality from lower to higher levels of distortion. However, it is observed that for some combinations of sources and codecs, there are some fluctuations that deviate from the expected monotonic trend, e.g., for AVIF sources \#2 and \#7. There are also some clear outliers, e.g. for VVC Intra source \#9, points that are far away from the fitted curve. 
These deviations indicate that while the general structure of the scale is captured, there are inconsistencies in the subjective rating. 

\par Several factors could contribute to this effect, such as the inherent difficulty of rating images with subtle differences in quality using this methodology, or the behavior of the image codec, that may distribute errors differently across the image (in some cases being more visible) when the distortion point changes. Additionally, some images exhibit negative scores. This implies that the corresponding distorted images were, counterintuitively, rated higher in quality than the original pristine images. This phenomenon can be attributed to the limitations inherent in the subjective assessment methodology for higher quality images, as well as the probabilistic scale reconstruction model, which does not restrict the DMOS to the $[0, 100]$ scale.

\subsection{DMOS Distribution Analysis}

In \cite{saupe2024maximum}, optimization methods for modeling ACR image and video quality rating distributions were investigated, achieving acceptable goodness of fit and outperforming empirical distributions in predicting unseen data. Similarly, the outlier removed scores (Section \ref{sec:outlier_removal}) for each distorted stimulus are represented using a Beta distribution, which is
highly flexible by taking on various shapes (e.g., uniform, unimodal, U-shaped) and is able to capture the diverse distributions of quality scores. The scores for each question were first normalized to the unit interval $[0,1]$ by dividing the scores by 100. The Beta distribution has two shape parameters, $\alpha, \beta > 0$, which were estimated using Maximum Likelihood Estimation (MLE), or the method of moments if MLE failed to converge. When $\alpha = \beta$, the distribution is symmetric, and U-shaped when $\alpha = \beta \le 1$.  Additionally, the suitability of the beta distribution was assessed using a Chi-Square goodness-of-fit test at a significance level of $0.05$. The Chi-Square test compares the observed frequencies of data points within predefined bins (categories) against the expected frequencies under the Beta distribution. The test evaluates whether the observed distribution significantly deviates from the expected Beta distribution. The results showed that $93\%$ of the distributions per stimulus passed the test. 

\par
\begin{figure}[t!] 
\centering \includegraphics[width=1\linewidth]{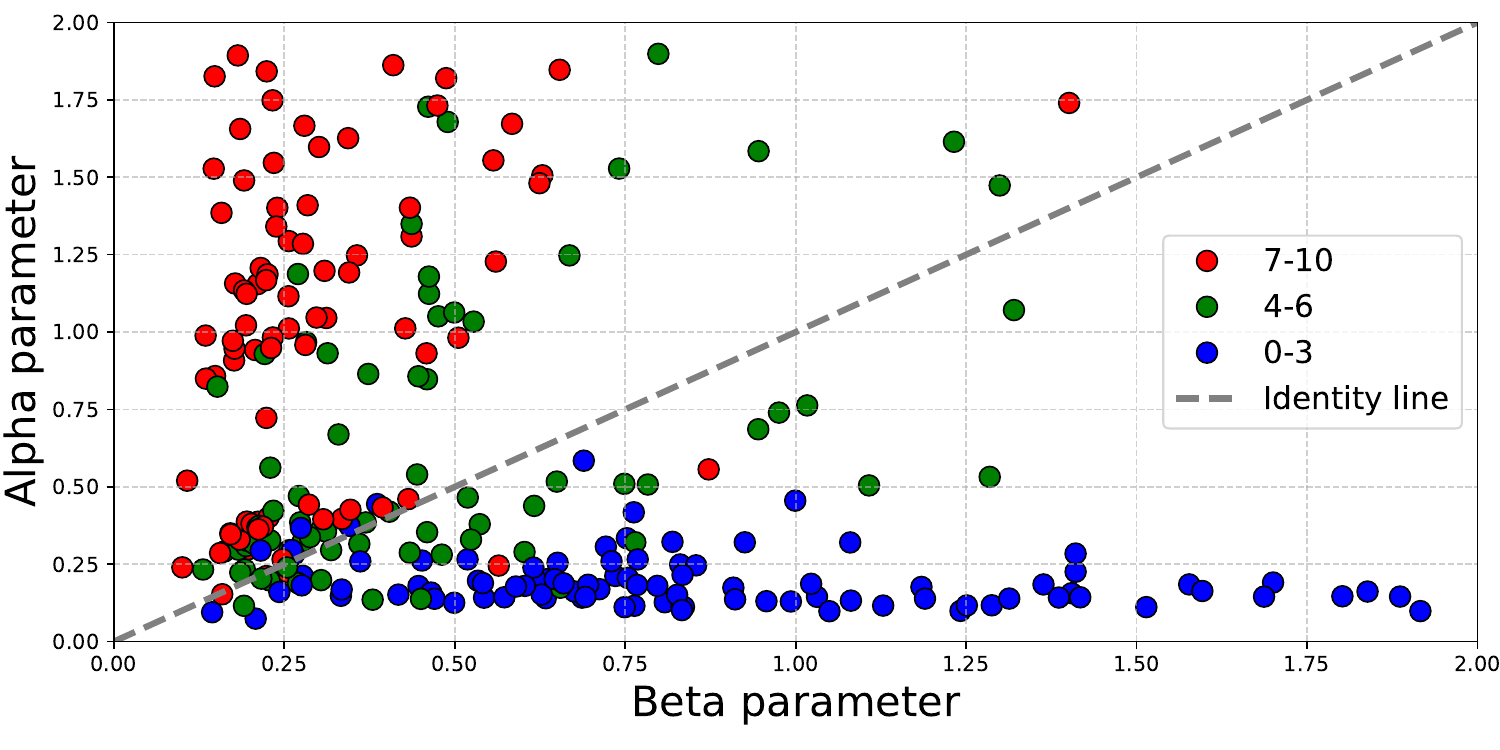}
\vspace{-25pt}
\caption{Beta distribution parameters for each question.} 
\label{fig:beta-distribution-parameters} 
  \vspace{-10pt}

\end{figure}

 The $\alpha$ and $\beta$ values associated to the distribution of each question are plotted in Fig. \ref{fig:beta-distribution-parameters}. When the $\alpha + \beta$ is high, the scores cluster around the mean (higher confidence) while when $\alpha + \beta$ is low, there is greater variability or disagreement among raters. Moreover, when $\alpha \approx \beta$, i.e. the point is close to the identity line ($y = x$), the quality scores are symmetric. As shown in Fig. \ref{fig:beta-distribution-parameters}, this happens more frequently for distortion levels 4, 5, and 6. This means that for these distortion levels, the ratings were not consistent among subjects and thus IDSQS methodology is not very reliable. In contrast, at the other distortion levels, the data points are concentrated near either the y-axis or the x-axis, indicating skewed distribution which is rather expected since these decisions are easier to make. For such cases, the distribution is uni-modal (one parameter dominates) and the peak of the distribution is close to 0 or 100 for the higher and lower distortion levels, respectively.

\subsection{JPEG AIC-3 Comparison}
In this section, the proposed double stimulus subjective assessment procedure is evaluated. As benchmark, the BTC-PTC methodology as described in Section \ref{sec:related_work} is used. These two procedures provide precise and accurate scores for a fine-grained quality scale as reference. Both of these methodologies, use triplet comparisons and BTC uses a boosting technique to enable a more fine discrimination between different qualities. 

\par
\begin{figure}[ht!]
    \centering
    \includegraphics[width=1\linewidth]{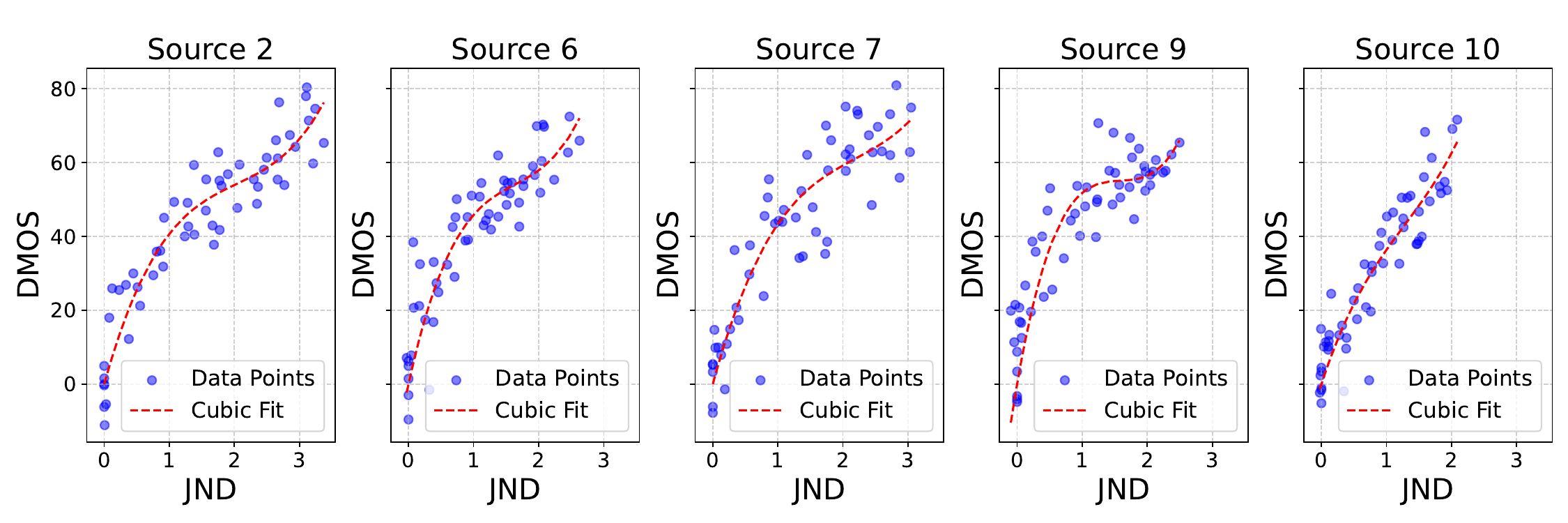}
    \vspace{-19pt}
    \caption{Performance comparison of IDSQS with JPEG AIC BTC-PTC \cite{testolina2025fine}.}
    \label{fig:comparison}
\end{figure}

 The reconstructed quality scores obtained from the data processing method described in Section \ref{sec:qual_recon} are compared to the reconstructed JNDs obtained for the BTC-PTC methodologies \cite{testolina2025fine}. This is a fair comparison since both reference and degraded images are the same. Therefore, the IDSQS performance evaluation presented in this section aims to assess the alignment between the DMOS reconstructed quality and the reference JND units. The results, aggregated by source content, are illustrated in Fig.\ref{fig:comparison}. For each source, a scatter plot of the reconstructed quality DMOS scores versus corresponding JND units is shown. To align the different scales, a cubic regression according to ITU-T P.1401 was employed to map DMOS to JND units. After the mapping, the Pearson, Spearman and Kendall's Tau between the perceptual quality indicators provided by JND and DMOS was also computed, to quantify the correlation between these two sets of scores. The results are shown in Table \ref{tbl-dmos-jnd}. 

\begin{table}[htb!]
\vspace{-14 pt}
\caption{Performance evaluation with correlation coefficients.}
\vspace{-5 pt}
\centering
\tiny
\resizebox{0.9\linewidth}{!}{
\begin{tabular}{l|c|c|c|c|c|c}
\hline
\specialrule{0.1em}{\abovetopsep}{\belowbottomsep}

\multirow{2}{*}{Method} & \multicolumn{6}{c}{Source images} \\
\cline{2-7}
& $2$ & $6$ & $7$ & $9$ & $10$  &$All$\\
\hline
\specialrule{0.1em}{\abovetopsep}{\belowbottomsep}

PLCC    &0.95 &0.91 &0.92 &0.91 &0.94   &0.89      \\
                           
\hline

SROCC    &0.93 &0.92 &0.89  &0.88  &0.94   &0.88    \\
                               
\hline

Kendall-tau   &0.78 &0.76 &0.72  &0.70  &0.80   &0.70     \\
                           
\hline
                  

\specialrule{0.1em}{\abovetopsep}{\belowbottomsep}
\end{tabular}
}
\vspace{-5pt}
\label{tbl-dmos-jnd}
\end{table}

As presented in Table \ref{tbl-dmos-jnd}, sources \#2 and \#10 have the highest PLCC between the reconstructed quality scales and the JND units. These are the source images where the proposed IDSQS methodology has the best performance, i.e. where the perceptual differences in quality are better captured. In contrast, image sources \#6 and \#9, display the lowest PLCC values, implying weaker alignment between the reconstructed scales and the JND units for these images. These discrepancies may arise from specific characteristics of the content, such as texture complexity or compression sensitivity. More specifically, both sources \#6 and \#9, depict natural scenes with regions that have high textural details, where artifacts can be hidden more easily. However, in the BTC-PTC methodologies, boosting techniques are used and thus, artifacts are more easily detected, even in these regions. When all the sources are considered, PLCC and SROCC are low which show that for this quality range IDSQS cannot estimate fine-grained quality scores and maintain consistent perceptual ordering, potentially due to content-specific variations or limitations in the subjective evaluation methodology.

\section{Conclusion} \label{sec:conclusions}
This paper proposes a double-stimuli subjective assessment methodology which uses in-place toggle between the reference and distorted image, allowing subjects to better identify artifacts. An extensive crowdsourcing study was performed, along with suitable methods to process the data. The experimental results have shown that despite the simplicity of the proposed double stimulus method, it can achieve good performance when compared to more reliable methodologies, such as those that are defined in JPEG AIC-3. Moreover, it has lower cost and thus is a practical alternative for DMOS estimation when fine-grained JND precision is not required.

\section*{Acknowledgment}
 The IST affiliated authors were supported by FCT/MECI through national funds and when applicable co-funded EU funds under UID/50008: Instituto de Telecomunicações. Mohsen Jenadeleh was funded by the Deutsche Forschungsgemeinschaft (DFG), Project ID 496858717. Dietmar Saupe was funded by the DFG, Project ID 251654672 – SFB TRR 161. The EPFL authors would like to thank the Swiss National Foundation for Scientific Research (SNSF) under grant number 200020 207918 for funding this research. 

\bibliographystyle{ieeetr}
\bibliography{paper}

\begin{thebibliography}{10}

\bibitem{wang2004image}
Z.~Wang, A.~C. Bovik, H.~R. Sheikh, and E.~P. Simoncelli, ``Image quality assessment: from error visibility to structural similarity,'' {\em IEEE Transactions on Image Processing}, vol.~13, no.~4, pp.~600--612, 2004.

\bibitem{BT.500}
{Recommendation ITU-T BT.500-15}, ``Methodologies for the subjective assessment of the quality of television images,'' 2023.

\bibitem{P.910}
{Recommendation ITU-T P.910}, ``Subjective video quality assessment methods for multimedia applications,'' 2023.

\bibitem{aic1}
{ISO/IEC TR 29170-1:2017}, ``{Information technology — Advanced image coding and evaluation — Part 1: Guidelines for image coding system evaluation}.''

\bibitem{aic2}
{ISO/IEC 29170-2:2015}, ``{Information technology — Advanced image coding and evaluation — Part 2: Evaluation procedure for nearly lossless coding}.''

\bibitem{BT1082}
{ITU-R BT.1082-1}, ``Studies toward the unification of picture assessment methodology,'' 1990.

\bibitem{ISO29170-32025}
{ISO/IEC DIS 29170-3}, ``{Information technology — JPEG AIC Assessment of image coding — Part 3: Subjective quality assessment of high-fidelity images},'' 2025.

\bibitem{testolina2023performance}
M.~Testolina, D.~Lazzarotto, R.~Rodrigues, S.~Mohammadi, J.~Ascenso, A.~M. Pinheiro, and T.~Ebrahimi, ``On the performance of subjective visual quality assessment protocols for nearly visually lossless image compression,'' in {\em Proc.\ 31st ACM International Conference on Multimedia}, pp.~6715--6723, 2023.

\bibitem{men2021subjective}
H.~Men, H.~Lin, M.~Jenadeleh, and D.~Saupe, ``Subjective image quality assessment with boosted triplet comparisons,'' {\em IEEE Access}, vol.~9, pp.~138939--138975, 2021.

\bibitem{testolina2025fine}
M.~Testolina, M.~Jenadeleh, S.~Mohammadi, S.~Su, J.~Ascenso, T.~Ebrahimi, J.~Sneyers, and D.~Saupe, ``Fine-grained subjective visual quality assessment for high-fidelity compressed images,'' in {\em 2025 Data Compression Conference (DCC)}, pp.~123--132, 2025.

\bibitem{jenadeleh2025subjective}
M.~Jenadeleh, J.~Sneyers, P.~Jia, S.~Mohammadi, J.~Ascenso, and D.~Saupe, ``Subjective visual quality assessment for high-fidelity learning-based image compression,'' {\em arXiv preprint arXiv:2504.06301}, 2025.

\bibitem{testolina2023jpeg}
M.~Testolina, V.~Hosu, M.~Jenadeleh, D.~Lazzarotto, D.~Saupe, and T.~Ebrahimi, ``{JPEG AIC-3 Dataset: Towards defining the high quality to nearly visually lossless quality range},'' in {\em 2023 15th International Conference on Quality of Multimedia Experience}, pp.~55--60, 2023.

\bibitem{naderi2024crowdsourcing}
B.~Naderi and R.~Cutler, ``A crowdsourcing approach to video quality assessment,'' in {\em IEEE International Conference on Acoustics, Speech and Signal Processing (ICASSP)}, pp.~2810--2814, IEEE, 2024.

\bibitem{Otsu}
N.~Otsu, ``A threshold selection method from gray-level histograms,'' {\em IEEE Transactions on Systems, Man, and Cybernetics}, vol.~9, no.~1, pp.~62--66, 1979.

\bibitem{saupe2024maximum}
D.~Saupe, K.~Rusek, D.~H{\"a}gele, D.~Weiskopf, and L.~Janowski, ``Maximum entropy and quantized metric models for absolute category ratings,'' {\em IEEE Signal Processing Letters}, vol.~31, pp.~2970--2974, 2024.

\end{thebibliography}

\end{document}